\def\nice#1::::{#1}    \def\subm#1::::{}   
\def\subm#1::::{}

\def\hMpc{\mbox{$h^{-1}\mbox{\rm Mpc}$}}
\def\centreline{\centerline}
\def\.{{\cdot}}
\def\gtapprox{\,\lower.6ex\hbox{$\buildrel >\over \sim$} \, }
\def\ltapprox{\,\lower.6ex\hbox{$\buildrel <\over \sim$} \, }

\def\etal{\mbox{et\,al.}}
\def\arcs{\ifmmode {'' }\else $'' $\fi}     
\def\arcm{\ifmmode {' }\else $' $\fi}     
\def\deg{\ifmmode^\circ\else$^\circ$\fi}    

\def\jref#1;;#2;;#3;;#4 {#1, {#2, }{#3, }#4}

\def\bref#1;;#2;;#3;;#4;;#5 {#1, {in #2 }(#3: #4), #5}
\def\apjpre#1;; {#1, preprint}

\def\apjpriv#1;; {#1, private communication}

\def\apjprpn#1;; {#1, in preparation}

\def\apjsub#1;;#2;; {#1, #2, submitted}

\def\apjprss#1;;#2;; {#1, #2, in press}

\nice \input epsf ::::

\documentstyle{mn}
\begin{document}

\def\IAP{Institut d'Astrophysique de Paris, 98bis Bd Arago, F-75.014 Paris,
France}
\def\sussex{Astronomy Centre, 
MAPS, University of Sussex, Falmer, Brighton,
BN1~9QH, United Kingdom}

\title[Topology of the 
Observable Universe]{On Determining the Topology of the 
Observable Universe via 3-D Quasar Positions}

\author[B.F.~Roukema]{B.F.~Roukema$^{1,2}$\\
{$^1$\IAP}\\$^2$\sussex }


\def\oo{$\ddot{\mbox{\rm o}}$}
\def\LaLu{Lachi\`eze-Rey \& Luminet (1995)}

\maketitle

\newfont{\sans}{cmss10}


\begin{abstract}
   Hot big bang cosmology says nothing about the topology of the Universe.
A topology-independent 
algorithm is presented
which is complementary to that of Lehoucq \etal\  1996 and
which searches for evidence of multi-connectedness using 
catalogues of astrophysically observed objects.    
The basis of this algorithm is simply to search for a quintuplet of 
quasars (over a region of a few hundred comoving Mpc) which can
be seen in two different parts of our past time cone, allowing for
a translation, an arbitrary rotation and possibly a reflection. 
This algorithm is demonstrated by application to the 
distribution of quasars between redshifts of $z=1$ and $z\approx4,$
i.e., at a comoving distance from the observer $1700 h^{-1}\mbox{\rm Mpc}
\ltapprox d \ltapprox 3300 h^{-1}\mbox{\rm Mpc}.$
Two pairs of isometric quintuplets separated by 
more than {$300$\hMpc} 
 are found. 
This is  consistent with the number expected from 
Monte Carlo simulations in a simply connected Universe
if the detailed anisotropy of sky coverage
by the individual quasar surveys is taken into account.
The linear transformation in (flat) comoving space from one quintuplet to
another requires translations of {$353$\hMpc}
and {4922\hMpc} respectively, plus a reflection in the former case, and 
plus both a rotation and a reflection in the latter.
Since reflections are required, 
if these two matches were due to multi-connectedness, then the Universe would 
be non-orientable.
\end{abstract}


\def\tabone{
\begin{table*} 
\caption{\label{t-2matches} Positions of the pairs of isometric quasar quintuplets}
$$\begin{array}{l c c c@{\ \ \ }l c c c}
\multicolumn{1}{c}{{\rm Object}} & \multicolumn{1}{c}{{}\alpha{}} & \multicolumn{1}{c}{{}\delta{}} &
\multicolumn{1}{c}{z} &
\multicolumn{1}{c}{{\rm Object}} & \multicolumn{1}{c}{{}\alpha{}} & \multicolumn{1}{c}{{}\delta{}} &
\multicolumn{1}{c}{z} \\
\hline 
\multicolumn{8}{c}{{\rm First\  Quintuplet\  Pair}}\\
\hline 
 {\rm [HB89]\ 0122-334 }      &  1{}^h{}22{}^m{}37.9{}^s{} & -33{}\deg{}25{}\arcm{}43{}\arcs{} &
 2.260 &
 {\rm  [HB89]\ 0103-271 }      &  1{}^h{}03{}^m{}07.8{}^s{} & -27{}\deg{}07{}\arcm{}37{}\arcs{} &
 2.180 \\
  {\rm [HB89]\ 0114-358  }     &  1{}^h{}14{}^m{}13.2{}^s{} & -35{}\deg{}52{}\arcm{}42{}\arcs{} &
 2.260 &
  {\rm [HB89]\ 0050-280}       &  0{}^h{}50{}^m{}05.8{}^s{} & -28{}\deg{}04{}\arcm{}24{}\arcs{} &
 2.152 \\
  {\rm [HB89]\ 0123-368 }      &  1{}^h{}23{}^m{}38.8{}^s{} & -36{}\deg{}48{}\arcm{}12{}\arcs{} &
 2.205 &
 {\rm  [HB89]\ 0049-261}       &  0{}^h{}49{}^m{}30.1{}^s{} & -26{}\deg{}06{}\arcm{}48{}\arcs{} &
 2.060 \\ 
 {\rm  [HB89]\ 0109-355}        &  1{}^h{}09{}^m{}07.5{}^s{} & -35{}\deg{}31{}\arcm{}23{}\arcs{} & 2.307 &
  {\rm [HB89]\ 0049-290 }      &  0{}^h{}49{}^m{}29.8{}^s{} & -29{}\deg{}00{}\arcm{}41{}\arcs{} & 2.220 \\ 
 {\rm [HB89]\ 0113-327\ NED02} & 1{}^h{}13{}^m{}30.8{}^s{} & -32{}\deg{}46{}\arcm{}23{}\arcs{} & 2.260 &
  {\rm [HB89]\ 0103-290   }    &  1{}^h{}03{}^m{}02.7{}^s{} & -29{}\deg{}05{}\arcm{}53{}\arcs{} &  2.230 \\
 \hline 
\multicolumn{8}{c}{{\rm Second\  Quintuplet\  Pair}}\\
\hline 
 {\rm  [CCS88]\ 133918.2+271659}&13{}^h{}39{}^m{}18.2{}^s{} &  27{}\deg{}16{}\arcm{}59{}\arcs{} &
 1.754 &
  {\rm [HB89]\ 0104-381\ NED02} &  1{}^h{}04{}^m{}56.2{}^s{} & -38{}\deg{}07{}\arcm{}07{}\arcs{} &
 2.060 \\
  {\rm [HB89]\ 1339+280  }     & 13{}^h{}39{}^m{}22.8{}^s{} &  28{}\deg{}02{}\arcm{}04{}\arcs{} &
 1.965 &
 {\rm  [HB89]\ 0117-374  }     &  1{}^h{}17{}^m{}40.4{}^s{} & -37{}\deg{}25{}\arcm{}29{}\arcs{} &
 1.940 \\
  {\rm [HB89]\ 1343+266\ NED01} & 13{}^h{}43{}^m{}24.7{}^s{} &  26{}\deg{}40{}\arcm{}06{}\arcs{} &
 2.030 &
 {\rm  [HB89]\ 0114-368  }     &  1{}^h{}14{}^m{}44.8{}^s{} & -36{}\deg{}52{}\arcm{}21{}\arcs{} &
 1.820 \\
 {\rm  [CCS88]\ 133211.4+280201} & 13{}^h{}32{}^m{}11.4{}^s{} &  28{}\deg{}02{}\arcm{}01{}\arcs{} &
 1.706 &
 {\rm  [HB89]\ 0106-379  }     &  1{}^h{}06{}^m{}54.7{}^s{} & -37{}\deg{}56{}\arcm{}18{}\arcs{} &
 2.200 \\
 {\rm  [CCS88]\ 133243.2+272958} & 13{}^h{}32{}^m{}43.2{}^s{} &  27{}\deg{}29{}\arcm{}58{}\arcs{} &
 1.515 &
  {\rm [HB89]\ 0055-387   }    &  0{}^h{}55{}^m{}01.8{}^s{} & -38{}\deg{}44{}\arcm{}44{}\arcs{} &
 2.350 \\
\hline
\end{array} $$
\end{table*}
} 

\def\Figmink{ 
\begin{figure} 
\nice 
\vspace{5cm} ::::                            
\subm \vspace{5cm} ::::
\caption{\label{f-LSSc} Example of ``pseudo-structure'' greater than
the size of a compact Universe. The image is an Aitoff all-sky 
projection of a sphere of horizon radius ($6000h^{-1}$~Mpc) 
in the covering space;
galactic latitudes smaller than 
20\deg are excluded. The grey-scale indicates
the value of $\delta T/T$ for each point on this sphere, where
$\delta T/T$ is defined 
within the fundamental cube of size ($1000h^{-1}$~Mpc$)^3$ 
by 
$\delta T(x,y,z)/T(x,y,z)= \sin(2\pi x/1000h^{-1}\mbox{\rm\,Mpc}).$ 
The axes of the fundamental cube are arbitrarily rotated by $0\.5$ radians 
from the $l^{II}=0\deg$ and $b^{II}=0\deg$ axes.
This all-sky map is not meant to be a realistic simulation of the CMB;
it merely demonstrates that a small Universe containing a 
perturbation in $\delta T/T$ can appear to have 
structure in the covering space 
which is much larger than the size of the Universe itself. Note that 
these structures should not be confused with the circles predicted 
by Cornish {\etal} 1996 for a multi-connected Universe: in the latter
case the circles are not (except in exceptional cases) of constant
temperature.
}
\end{figure} 
}

\def\FigOne{ 
\begin{figure}  
\nice \centreline{\epsfxsize=8cm
\epsfbox[82 75 480 450]{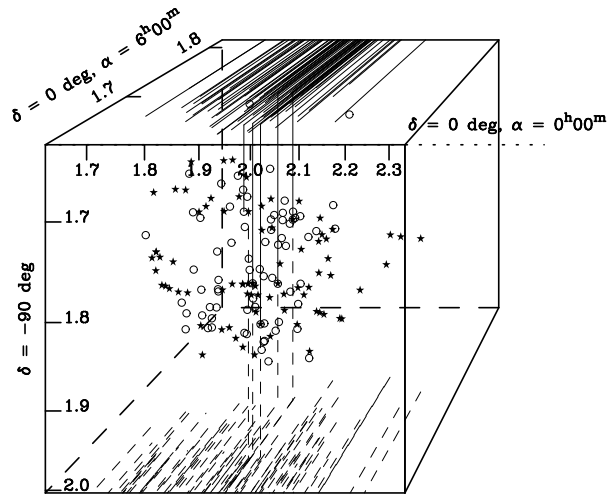} } ::::
\subm \vspace{9cm} ::::
\caption{\protect\footnotesize
Superposition of quasars within {200\hMpc} of [HB]~0122-334 
with the quasars within {200\hMpc}
 of [HB89]~0103-271, using the linear transformation 
mentioned in the text. 
(All distances mentioned here are comoving.) 
The axes are parallel to the directions in right ascension and
declination as labelled, units are in redshift. Unshifted quasars are shown as 
stars, shifted quasars are shown as open circles. For clarity, vertical lines 
are omitted except for the members of the matching quintuplets.
}
\end{figure} 
}



\def\FigTwo{ 
\begin{figure}  
\nice \centreline{\epsfxsize=8cm
\epsfbox[82 75 480 450]{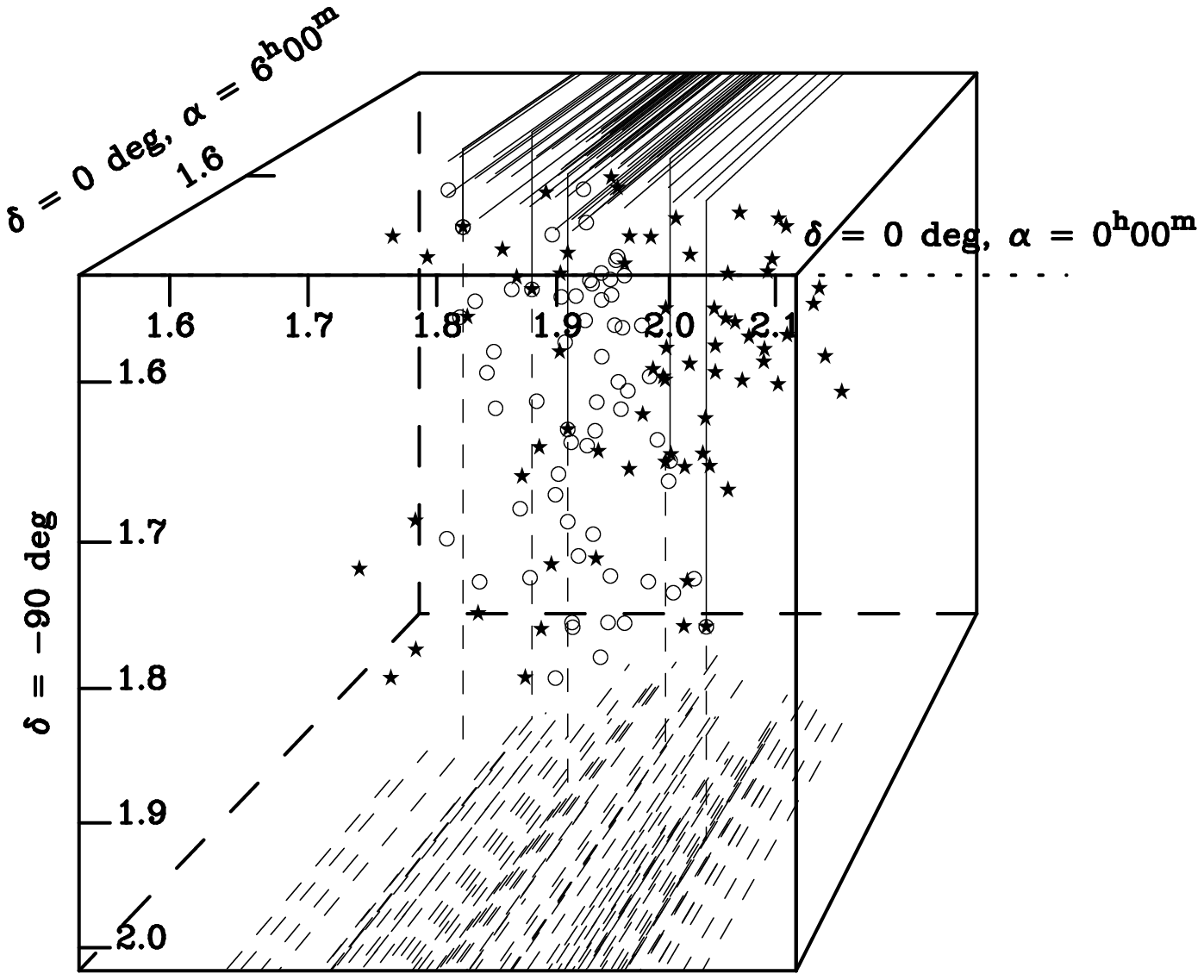} } ::::
\subm \vspace{9cm} ::::
\caption{\protect\footnotesize
Superposition of quasars within {200\protect\hMpc} of  [CCS88]~133918.2+2716
with the quasars within {200\hMpc} of  [HB89]~0104-381~NED02, using the linear transformation 
mentioned in the text. Axes and symbols are as for Fig.~1. 
}
\end{figure} 
}

\section{Introduction}

	Given the immense observational effort to search for 
evidence of the possible non-trivial
curvature of the observable Universe, (in particular for an ``open'', i.e.,
negatively curved, $\Omega_0<1$, Universe) it is surely 
equally important to search for evidence
for that the observable Universe could have a non-trivial topology. 
Apart from the obviously fundamental significance
of knowing that the volume of the Universe is finite
(if the topology turns out to be compact),
studies of evolution of 
galaxies, quasars and other objects at cosmological distances would be 
immensely advanced by the ability to observe the same object at 
significantly different epochs. 
General relativity and the standard hot big bang model of the Universe
are independent of topology.
According to several authors (e.g., Hawking 1984; 
Zel'dovich \& Grishchuk 1984),
quantum gravity with a ``no-boundary'' boundary condition requires 
the Universe to be compact. 
A multi-connected topology of the observable 
Universe would provide such compactness.

Lachi\`eze-Rey \& Luminet (1995) present an extensive review of 
what multi-connectedness of our Universe would mean 
and 
of ef\-forts made so far to de\-tect multi\--connected\-ness, 
so only a 
short introduction is given here. 

A simple example of a 3-dimensional manifold with a non-trivial
topology is the hypertorus. This can be constructed by identifying
opposite faces of a cube (or more generally, a rectangular prism). 
A particle which one would otherwise expect 
to leave the cube at one surface ``reenters'' at the opposite face, 
without ``noticing'' that anything strange has happened. 
Many N-body simulations used to calculate the non-linear effects of 
gravity use a hypertoroidal topology in order to provide boundary 
conditions; this is termed using ``periodic'' boundary conditions.
If the universe were hypertoroidal, 
then the identification of opposite faces 
would be physical, not merely a numerical technique. 

A useful way of thinking about this in terms
of a simply connected Universe is to
imagine the cube repeated endlessly in a 3-dimensional grid. 
Each repetition consists
of the same physical region of space, and photon paths (geodesics) 
can be calculated
just as for the simply connected Universe. Our past time cone can then
be thought of as before---but 
with the difference that we see the same chunk of the
(whole) Universe at earlier and earlier epochs further and further away.
This apparent space containing multiple copies of the Universe is 
called a ``covering space''. In the case of the hypertorus, and if 
objects (e.g., quasars) had visible lifetimes as great as that of the Universe
and were all bright enough to be seen in magnitude/surface-brightness  
limited surveys,
then an obvious grid pattern would appear. 

	However, the hypertorus is only one of many possible topologies. 
There is an immense variety
of topologies of the many possible 3-manifolds. For example, there are ten
flat, homogeneous, isotropic, multi-connected 3-manifolds of finite
volume (``compact''), and the classification of negatively curved,
homogeneous, isotropic, multi-connected, compact 3-manifolds 
is still an open area of research (e.g., Thurston \& Weeks 1984).
Because of evolutionary effects and the vast range of possible topologies,
the grid pattern possible for a hypertorus would not
necessarily be obvious to the eye, so subjective impressions of present
data are certainly not sufficient to rule out a non-trivial topology.

    Other topologies can be thought of by starting with a ``fundamental
polyhedron'' and identifying pairs of faces. Using the conventions of
\LaLu, the shortest distance from an object to any of its ``ghost'' 
counterparts is 
labelled $\alpha$, and the largest distance from an object to an 
adjacent ghost (where there is an adjacent ghost for each 
face of the fundamental
polyhedron) is labelled $\beta$. The choice of a fundamental 
polyhedron for thinking about a multi-connected topology is not
unique; such a Universe (as considered here) is on average homogeneous and
isotropic, without any borders in hypersurfaces of 
constant cosmological time.

	The claimed periodicity of the galaxy redshift distribution 
in pencil-beam surveys 
(Broadhurst et al. 1990) could have been an indication of a non-trivial
topology, but closer investigation of the data has not confirmed this.
As argued thoroughly by \LaLu, observations on scales larger
than this are not yet extensive enough to rule out repetitions 
of a finite-volume multiconnected Universe above
a scale of about $\beta=600 h^{-1}$~Mpc
($\alpha=300 h^{-1}$~Mpc). 

\Figmink 

	Intuitively, one might think that a simple way to test for
a compact topology would be to note 
that in a Universe of finite size no objects could exist which are
larger than the size of the Universe, so that the initial perturbation 
spectrum (e.g., as observed by COBE) should be zero at large length 
scales, or small wavenumbers $k.$ The flaw in this argument is
that in the covering space, structures which physically occupy
the same, or nearly the same, part of space can coincide at the
``faces'' of the fundamental polyhedron in such a way 
that a pseudo-structure which is larger than the Universe can exist
in the covering space. In analysis which works in the covering
space (which is much easier than starting with a hypothesised
topology), such a pseudo-structure would violate the minimum-$k$
criterion. As a demonstration, Fig.~1 shows a simple example of 
a sky map due to such pseudo-structure
in a hypertoroidal Universe which 
is only a sixth of the horizon size. The huge structures
which extend across the sky are simply due to the ``microwave 
background'' sphere cutting through the same (physical) 
structure at slightly different angles in adjacent ghost
images of the fundamental polyhedron (cube). 

	While such a dramatic pattern should be easy to rule out,
it is not clear what 
assumptions would be needed  about the initial perturbation
spectrum and the topology in order to 
make the minimum-$k$ test useful.
In the example here, the maximum size of the fundamental polyhedron 
is clearly visible in the direction
perpendicular to that of the pseudo-structures, so that 
averaging over the two dimensions of the sphere should help
in the derivation of a generalised minimum-$k$ test. However, without a
rigorous derivation of such a test, 
detailed models with assumptions about structure formation and
specific topologies are required.
Some authors have indeed calculated detailed models for particular
cosmologies. In particular, toroidal topologies have been 
shown to be inconsistent with the COBE data to the scales of
several thousand $h^{-1}$~Mpc (Stevens et al. 1993)
to around the horizon size 
(Starobinsky 1993; de Oliveira \& Smoot 1995)
on the basis of considering a few (toroidal) of the topologies possible.
Others (Jing \& Fang 1994) note that the COBE data is well fit by 
a non-zero low wavelength cutoff in the 
primordial fluctuation spectrum, which would be consistent with a 
finite-volume, multiconnected Universe at about $1-3$ times 
the horizon scale. 
	
	An alternative to a minimum-$k$ test modified in some way by 
prior assumptions is the 
topology-independent method for searching for
multi-connectedness in present and future CMB data which 
has been proposed by 
Cornish {\etal} (1996). This method is based on the property 
that a fundamental polyhedron of about the horizon
size or smaller should intersect the last scattering sphere several
times, in circles.
Different copies of an intersection which 
represent the same subset of space-time should be seen in different
parts of a CMB all-sky map. As long as the temperature of a point 
in the CMB is independent of the direction of the observer (e.g.,
the ``Doppler'' fraction 
of $\delta T/T$ due to the movement of the baryon-photon
fluid towards or away from the observer is small), the values of
 $\delta T/T$ around corresponding circles should be identical apart
from a rotation due to the particular multi-connected manifold 
explaining the identification. Note that the circles themselves 
are not of constant temperature: the pseudo-structures shown in
Fig.~1 are not the same as the circles discussed by Cornish {\etal}.

	To put these tests into perspective, the reader
is reminded that in comoving units, i.e., 
within the hypersurface of constant $t=t_0,$ 
the distance to an object at redshift $z$ in 
the hypersurface of constant $t=t_0$ is $d= 2(c/H_0) (1-1/\sqrt{1+z}),$
the horizon is at $6000 h^{-1}$~Mpc. 
(Except where otherwise stated, all distances here are quoted 
in comoving units.) 
 The diameter across the
observable Universe is, of course, twice the radius, i.e., $12000 h^{-1}$~Mpc,
so it could only be claimed that the observable Universe is
simply connected if evidence were available up to a scale of twice
the horizon size.

\section{Method}
   Lehoucq et al.'s (1996) result that $\alpha \gtapprox 300 h^{-1}$~Mpc
was obtained from galaxy clusters using a very simple and powerful 
search method which avoids having to make prior 
assumptions on what the topology is.
Their method consists of 
plotting a histogram of the apparent comoving separations 
between every pair of objects
in a catalogue. If the catalogue samples a covering space containing several 
ghost images of the Universe, then the apparent separations of 
identical objects in the covering space would correspond to
the sizes of various multiples and linear combinations of the translation
vectors (between one copy of the Universe and the next). 
For every pair of copies of the Universe (containing $N$ objects 
of a certain class), there will $N(N-1)$ apparent separations of 
non-identical objects will be scattered around
these ``translation vector'' values. Since the separations of identical
objects are not scattered, sharp peaks in the histogram will appear at
several of these favoured separations.

The most obvious candidates for detecting mul\-ti\--con\-nect\-ed\-ness on 
larger scales are quasars. 
However, applying Lehoucq \etal's method to the quasar population
would be weakened by evolutionary effects. 
Since
quasar lifetimes are probably an order of magnitude or more less than
the age of the Universe (e.g., Cavaliere \& Padovani 1988), 
many of the pairs of quasars separated by 
the linear combinations of ``translation vectors'' will be observed
at sufficiently different redshifts that 
they will only be bright enough to be visible at one of the two
epochs. With the additional problem that the part of our past time
cone systematically searched for quasars is still a small fraction of
that available, the repetition effect necessary for obtaining
sharp peaks is likely to be weak.

	Therefore, I have chosen a more direct, (but slightly harder to 
program), algorithm which, as for Lehoucq et al.'s (1996) and 
Cornish \etal's (1996) methods, 
does not assume any particular choice for the topology.

	The idea is to search for a repetition in the 
universal covering space by searching
for a quintuplet of several ``close''
quasars which exist in the same relative
geometrical relation, modulo a rotation and/or a reflection, in at
least two different parts of the observed universal covering space
(i.e., in the hypersurface at constant cosmological time $t=t_0$ 
which corresponds to the interpretation of observations in ignorance
of any possible multi-connectedness). 
The precise definition of ``close'' is that four of the quasars are each
within  $d_c \ltapprox 200h^{-1} $~Mpc of the fifth (arbitrarily chosen) 
quasar.   
Among the data set used (essentially that of quasars with $z>1$), 
containing $N=5114$ objects, the mean number of objects within this
radius is $M=21.$
There are $\sim N(N-1)/2$ pairs of quasars. For each of the two quasars in
a pair, there are $\sim M!/(M-4)!$ quintuplets which can be formed 
with four ``close'' companions.
A na\"{\i}ve implementation of this technique would therefore require
$\sim N(N-1)/2\; [M!/(M-4)!]^2 \sim N^2M^8 \ltapprox 10^{18}$ comparisons of
geometry between quintuplets, way beyond present computing capabilities.

	This number of operations can be reduced drastically by 
using subsets of the geometrical criteria 
to filter out impossible cases at appropriate stages of the algorithm, 
therefore wasting no more operations on such cases. The algorithm is as
follows. 

For a given pair of quasars, a list of the ``close'' companions 
(hereafter ``secondaries'') of each is
made. The distance of each secondary from its primary is calculated.
The lists of primary-secondary distances for the two primaries 
are searched for identical values. In the case of 
genuinely matching quintuplets (where the primaries correspond
in this match), the four primary-secondary
distances around each primary are identical, so should pass this first filter.
Hereafter, this filtering process is termed ``distance matching''. This ends
up with a list of zero, four or more candidate secondaries for each primary. 

For each list of remaining candidate secondaries, 
the angle formed by each pair of 
candidates with its primary quasar is calculated, making two lists 
of companion-primary-companion angles. 
These two lists of companion-primary-companion angles are then searched for 
identical values (between the two lists).
Hereafter, this is referred to as ``angle matching''. 
Once again, if there is a genuine
quintuplet match, this will create six companion-primary-companion angles
which match.
These two filters, ``distance matching'' and ``angle matching'', 
drastically reduce the number of 
secondaries remaining in the list of ``candidate'' secondaries, making 
the problem computable in practice.

Still considering the same pair of primary quasars,
a search is then made for four secondaries among which each of the six
possible secondary-primary-secondary angles has 
corresponding matches around
the second primary quasar. This guarantees that the two quintuplets
(each composed of primary plus four secondaries) are isometric 
within the tolerances chosen. 

	The choice of quintuplets rather than, say,  quadruplets or sextuplets,
is a compromise based on the data and computing power available. 
The probability of finding corresponding $n$-tuplets 
which are simply due to chance is larger for smaller $n,$ 
while the number of computing operations required is larger for 
larger $n.$

	The main astrophysical uncertainties which could affect this method 
(apart from the problem of the physics of quasar lifetimes!) are those in the
comoving radial distances. These are due to
(1) the uncertainty in the spectroscopic measurement of the
redshift and (2) the peculiar velocity of the quasar relative to the
comoving reference frame. The latter causes (2a) a movement relative
to the comoving reference frame between the two redshifts at which
the quasar is observed and (2b) a misestimate of the quasar's comoving radial 
distance due to inclusion of the peculiar velocity doppler shift as
part of the cosmological redshift.

The most serious of these constraints is (2b): a quasar at $z=2$ having
a peculiar radial velocity of $v_{pec}=400 \mbox{\rm km s}^{-1}$ has
its comoving radial distance misestimated by about $5h^{-1}$~Mpc. 
The ``movement'' error (2a) is likely to be less than about $0\.5 h^{-1}$~Mpc
(for $v_{pec}=400 \mbox{\rm km s}^{-1}$ and observations of the two
images at $z=1$ and $z=3$ the error is $0\.6 h^{-1}$~Mpc), 
while the measurement error (1) lies at either
of these two values depending on the whether the redshifts are available
to $\delta(z)=0\.01$ or $\delta(z)=0\.001$ respectively. (These estimates
assume $\Omega_0=1, \lambda_0=0$.)

%

\section{Results}
	This algorithm was applied to a catalogue of $N=5114$ objects 
(nearly all quasars,
but including a few galaxies) 
with redshifts greater than unity found in 
the NASA/IPAC Extragalactic Database (NED, 
primarily from surveys by Hewitt \& Burbidge (1989), 
Crampton et al. (1988), Boyle et al. (1990) and Schneider et al. (1992)).

	The distance defining ``close'' used was $d_c=$ $200h^{-1}$~Mpc, the
tolerance for primary-secondary distances to be considered identical
was $\Delta d=0.5${\hMpc} or $\Delta d=1$\hMpc, and the tolerance for 
secondary-primary-secondary angles to be considered identical
was $\Delta \theta=0\.01$ rad. The algorithm 
takes about $10^1$ hours to run on a Sun Sparcstation 10.

  Early attempts (using the $N=4848$ quasars available from NED in Oct 1994,
and  $\Delta d=0.5$\hMpc) 
resulted in 56  pairs of such quintuplets found in the data. 
Intuitively, it seems extraordinary that these match\-es could be due
to chance. However, a random 
data set was simulated, 
containing the same number of objects as in the quasar list
and distributed according to the same density distribution. As the
density distribution is not very uniform, combining a variety of different
observational selection criteria, this was done by retaining the celestial 
positions of all the objects in the observed data, but reassigning the
observed redshifts randomly to obtain a 3-dimensional random 
catalogue. This process should destroy any real matching quintuplets 
in the data. The results over several (8) simulations were 
$57\pm10$ ($1-\sigma$ uncertainty) pairs of matching quintuplets. 
So, if there are any real pairs of matching quasar quintuplets 
due to multi-connectedness, these
are swamped by 
coincidences simply due to Poisson statistics. 

	One way of trying to avoid the random coincidences 
was by introducing an independent criterion, 
the result of Lehoucq {\etal} (1996): 
the ``minimum'' size of the fundamental polyhedron (if it exists) 
must be greater than $\alpha=300$\hMpc. 

   Applying this requirement that the 
separation between the quintuplets is greater than 
$\alpha=300${\hMpc}  
(using the full $N=5114$ quasar catalogue and $\Delta d=1$\hMpc), 2 pairs of matching
quintuplets were found in the data, while only 
  $0\.54\pm0\.66$ were found in simulations (performed as before) 
based on this data.
Table~1 
lists positions (right ascension and declination, equinox  
B1950.0, and redshift) for the matching quintuplets. 
In Figures 1 and 2, the objects near a quintuplet are shown 
linearly transformed 
to the position of the ``ghost'' quintuplet. 
The transformations require 
translations of {353\hMpc} and {4922\hMpc} respectively, 
plus a reflection in the former case, and both a rotation and a reflection
in the latter case.
Since reflections are involved in both transformations, 
if the matches were due to multi-connectedness, then 
the Universe would be non-orientable. 

\tabone

\FigOne

\FigTwo 

	However, a more 
conventional explanation (within a simply-connected Universe) 
is available by noting 
that the Monte Carlo simulations could lose some clustering 
properties existent in the observational quasar distribution. 
Regions in which the (observed) quasar number density is higher 
have more objects, and therefore more chance correspondences.
Three of the twenty individual members of the matching quintuplets are 
in fact binaries (indicated by ``NED01'' or ``NED02'' in Table~1),
though this may not have any relation to clustering.
The underlying astrophyical clustering 
of quasars 
(after correction for selection effects) 
seems to follow galaxy clustering with a clustering
length of around 6-10\hMpc\ (Andreana \& Cristiani 1992), and so 
there should be 
little clustering at the scales of the quintuplets: 
separations between individual members of a quintuplet
range from 57\hMpc\  to 353\hMpc, with a median separation of 152\hMpc.

	Instead, an explanation lies at the level of ``observational'' 
clustering. The quasar sample is composed of a large 
number of individual surveys. In the creation of the random
simulations, the redistribution of the redshifts
of the entire catalogue therefore gives the same average redshift distribution
to each of the individual surveys. This average redshift distribution
may be sufficiently different from the original 
redshift distribution in a survey to affect the number of chance
coincidences. The full catalogue was therefore divided up into 93 
individual ``surveys'' over small solid angles such that 
the redshift distributions in individual catalogues were isotropic
within each ``surveyed'' region; the remaining 1751 quasars spread across the
sky were considered as a single ``survey''. Random simulations were
then performed by redistributing the redshifts within the individual
``surveys'', again retaining the observed set of positions on the sky.
The result over 15 simulations was $1\.1\pm1\.2$ isometric quintuplet
pairs, so the detection of 2 quintuplet pairs in the data is not
at all statistically significant. It serves instead as an illustration
of what a positive detection could be like.

\section{Conclusion}

	A topology-independent method,
complementary to those of Lehoucq {\etal} 1996 and Cornish {\etal} 1996, 
which observationally searches for the effects of a non-trivial topology
of the observable Universe has been presented. 
The method 
is simply to search for isometries of quasar quintuplets in different
parts of our past time cone, and an algorithm which reduces the
number of computations required to within practical computing
cabilities has been explained.

 This method is illustrated
by application to the observed set of quasars available from NED, 
and the significance is tested by comparison to Monte Carlo simulations.
Even by taking advantage of Lehoucq \etal's (1996) (and others') results 
against the Universe having a multi-connected topology on a scale 
$\alpha \le 300$\hMpc, the number of detections remains statistically
consistent with that expected for a simply connected Universe.

	Without a detailed physical model of quasar evolution,
the method is only capable of finding evidence for 
multi-connectedness, not for showing simple-connectedness. 
If a statistically significant number of corresponding quintuplets 
were found in a future more complete catalogue of quasars,
it should be followed by finding a fundamental polyhedron
and linear transformations which 
explain the 
quintuplet matches found. These should be applied to 
the full quasar catalogue in order to find the transformed
(``ghost'') quasar positions in the copy of the fundamental
polyhedron in which galaxy and galaxy cluster positions are 
best known, i.e., in the region of the covering space containing
the Galaxy. If the transformed positions of the quasars significantly 
corresponded
to the centres of, say, bright elliptical galaxies, or cluster
centres, this would be a confirmation of multi-connectedness,
and an indication of the evolutionary link between quasars and
galaxies. On the contrary, a limit on the absence of luminous 
objects at the transformed quasar positions would either be 
a refutation of the claimed multi-connected manifold or evidence
that the luminosity of ``dead'' quasars falls below the detection
limit.

\section{acknowledgements}
Thanks to Andrew Liddle, John Barrow and Amr El-Zant 
of the University of Sussex
Astronomy Centre for useful suggestions and comments. 
  This research has been carried out with support from the 
Institut d'Astrophysique de Paris, CNRS, from a PPARC
Research Fellowship and has 
made use of Starlink computing resources and 
the NASA/IPAC extragalactic database (NED)
which is operated by the Jet Propulsion Laboratory, Caltech, under contract
with the National Aeronautics and Space Administration. 



\end{document}